\documentclass[prb,twocolumn,amsmath,floats,showpacs]{revtex4}
\usepackage{graphicx}
\usepackage{textcomp}

\begin{document}
\input epsf

\title{Point-contact spectroscopy of electron-phonon interaction in superconductors}

\author{N.~L.~Bobrov \footnote {Email address: bobrov@ilt.kharkov.ua}, A.~V.~Khotkevich,
 G.~V.~Kamarchuk, and \fbox{P.~N.~Chubov}}
\affiliation{B.I.~Verkin Institute for Low Temperature Physics and
Engineering, of the National Academy of Sciences
of Ukraine, prospekt Lenina, 47, Kharkov 61103, Ukraine}

\published {Fiz. Nizk. Temp. 2014, \textbf{40}, p.280(Low Temp. Phys. 2014, \textbf{40}, p.215)}
\date{\today}

\begin{abstract}The possibility to reconstruct the electron-phonon interaction (EPI) function
was demonstrated for $S-c-N$ and $S-c-S$ point contacts using the superconducting inelastic
contribution to the excess current caused by Andreev reflection processes.
Superconductors with both weak (Sn, Al) and strong (Pb, In) EPI were considered.
It was shown that in the latter case it is necessary to account for the elastic component
of current which is related to the frequency dependence of the superconducting energy gap
arising due to electron-phonon renormalization of the energy spectrum of the superconductor.

\end{abstract}

\maketitle

\section{Introduction}

The significant progress recently made in the development
of new superconducting materials stirred a growing interest
in determining their key parameters. One of the most
important characteristics that allow analyzing the superconductor
behavior for the synthesis of new superconducting
compounds is the electron-phonon interaction (EPI) function.
One of the major techniques employed to determine the
EPI function is Yanson's point-contact (PC) spectroscopy.
This method has worked well for measuring nonlinearities of
current-voltage characteristics (CVC) in the point contacts
consisting of metals and compounds in the normal state\cite{Naidyuk1, Khotkevich1}.
On the other hand, for many superconductors it is very difficult
to realize the ballistic regime of current flow during their
transition to normal state, which is necessary for providing
the spectral mode of operation of a point-contact in
Yanson's point-contact spectroscopy. This problem is most
pronounced when the compounds belonging to the new
classes of superconducting materials are studied. The pointcontact
studies of such materials often have to deal with a
much distorted surface layer, which limits the possibility of
determining the EPI parameters. An effective solution to this
problem is to use the point-contact characteristics measured
in the superconducting state \cite{Bobrov}.

\section{Basic theoretical conceptions}

\subsection{Inelastic contribution}
The theory of inelastic spectroscopy of the EPI in a
superconductor considers ballistic point contacts with the
dimensions $d $ smaller than all characteristic lengths\cite{Khlus1, Khlus2, Khlus3}:
$d\ll \xi (0)$, $l_i$, $v_F$/$\omega_D$, where $\xi (0)$ is the
superconducting coherence length, $l_i$ is the scattering length on impurities,
$l_{\varepsilon} \sim {v_{F} } \mathord{\left/ {\vphantom {{v_{F} } {\omega_{D} }}}
\right. \kern-\nulldelimiterspace} {\omega_{D} }$ is the energy mean free path at
the Debye energy.

We should emphasize an important point: despite the
fact that most of the nonequilibrium phonons are generated
in the banks of the contact and any scattering process of the
Andreev electrons on nonequilibrium phonons is effective,
the existing theories consider only scattering in the region of
maximum concentration of nonequilibrium phonons since
the probability of their reabsorption by electrons depends on
the concentration of phonons. This region corresponds to the
highest current density and has a size, as also in the case of
Yanson's point-contact spectroscopy, of the order of the contact
diameter \cite{Kulik}.

The first publication\cite{Bobrov} on the reconstruction of EPI functions
from the spectra of superconducting point contacts has
addressed the cases that, to a certain extent, go beyond the
predictions of the theory of inelastic spectroscopy of EPI in
superconductors\cite{Khlus1, Khlus2, Khlus3}. In these cases, scattering in the banks
played an important role in the formation of nonlinearities in
such point contacts. However, the contacts which satisfy the
theoretical model to the fullest extent have not been considered.
In this paper, we will fill this gap and also consider the
point contacts in which the elastic contribution to electronphonon
scattering should be taken into account.

At the heart of the inelastic point contact spectroscopy
of superconductors lies the study of nonlinear current-voltage
characteristics of the contacts arising due to the
inelastic scattering of nonequilibrium phonons on electrons
undergoing Andreev reflection.

In Yanson's point contact spectroscopy the EPI function is
\vspace{5pt}

$G_{pc} \left( {eV} \right)=-\frac{3R_{0} \hbar v_{F} }{32ed}\cdot \frac{d^{2}I}{dV^{2}}$,
\vspace{5pt}

i.e., it is proportional to the second derivative of the currentvoltage
characteristic.\cite{Kulik} At the same time in the inelastic PC
spectroscopy of superconductors,\cite{Khlus1, Khlus2, Khlus3} the EPI function is
proportional to the first derivative of the excess current (the difference
between the current-voltage characteristics in the
normal and superconducting states at the same voltage). For
$S-c-S$ contacts, the following expression has been obtained\cite{Khlus1}:

\begin{equation} \label{eq__1}
\begin{array}{l} {\frac{dI_{exc} }{dV} =-\frac{64}{3R} \left(\frac{\Delta L}{\hbar \bar{v}} \right)\left[G^{N} (\omega )+\frac{1}{4} G^{S} (\omega )\right]} \\ {\quad \quad \quad \quad \quad \quad \quad \quad \quad \quad \quad \quad \quad \quad \quad \quad \omega ={eV\mathord{\left/ {\vphantom {eV \hbar }} \right. \kern-\nulldelimiterspace} \hbar } } \end{array}
\end{equation}

$G^{N} (\omega )$ is the PC EPI function identical to that of the point contact in the normal
state, $G^{S} (\omega )$ is the superconducting PC
EPI function different from $G^{N} (\omega )$ by a form factor. In contrast
to the normal form factor, which determines the contribution
to the current due to electron-phonon collisions
accompanied by a change in the $z$-component of the electron
velocity, in the case of the superconducting form factor
which is included in $G^{S} (\omega )$, it is the electron-phonon
collisions associated with Andreev reflection processes in
the contact region, i.e., conversion of quasi-electron into
quasi-hole excitations, that contribute to the current. The relative
magnitude of the phonon contribution to the excess
current is of the order of
${d\cdot \omega _{D} \mathord{\left/ {\vphantom {d\cdot \omega _{D}  v_{F} }} \right. \kern-\nulldelimiterspace} v_{F} } $,
for $eV\sim \omega _{D} $, i.e., it is small
if the condition
$d\ll {v_{F} \mathord{\left/ {\vphantom {v_{F}  \omega _{D} }} \right. \kern-\nulldelimiterspace} \omega _{D} } $
is fulfilled.

An analogous expression for $S-c-N$ contacts is \cite{Khlus2}

\begin{equation}\begin{split}
\label{eq__2}
\left. {\frac{1}{R(V)}- {\frac{1}{R(V)}} } \right|_{\Delta =0}
= \\ = -\frac{32d\Delta }{3R\hbar }\left[ {\frac{1}{v_{F}^{(1)} }G_{1} \left(
\omega \right)+ \frac{1}{v_{F}^{(2)} }G_{2} \left( \omega \right)} \right]
\end{split}\end{equation}

For the second derivative of the CVC in $S-c-N$ point contacts,
the following expression has been obtained:

\begin{equation}
\label{eq__3}
\frac{1}{R} \frac{dR}{dV}=\frac{16ed}{3\hbar }
\sum\limits_{a=1,2} {\frac{1}{v_{F}^{(a)} } \int\limits_0^\infty
{\frac{d\omega }{\Delta } S\left( {\frac{\omega -eV}{\Delta }}
\right)G_{a} (\omega )} }
\end{equation}

$G_{a} (\omega )$ are the EPI functions for the normal and superconducting
metals forming the heterojunction, $S(x)$ is the smearing factor,

\begin{equation}
\label{eq__4}
S(x)=\theta (x-1)\frac{2\left( {x-\sqrt {x^{2}-1} } \right)^{2}}{\sqrt
{x^{2}-1} },
\end{equation}

where $\theta (x)$ is the Heaviside theta-function. Thus, for $T\to 0$,
the resolution is determined by the value of $\Delta $.
From expression \eqref{eq__3}, given the relation between the derivative
of CVC and the PC EPI function, it can be written as

\begin{equation}
\label{eq__5}
\tilde{{g}}_{pc}^{S} =\int\limits_0^\infty {\frac{d\omega }{\Delta }S\left(
{\frac{\omega -eV}{\Delta }} \right)g_{pc}^{N} (\omega )}
\end{equation}

As a model $g_{pc}^{N} (\omega )$, we will take the EPI function of Cu-Sn
heterojunction reconstructed from its spectrum in the normal
state.

The calculation results obtained using \eqref{eq__5} in Fig.~1.
\begin{figure}[t]
\includegraphics[width=8cm,angle=0]{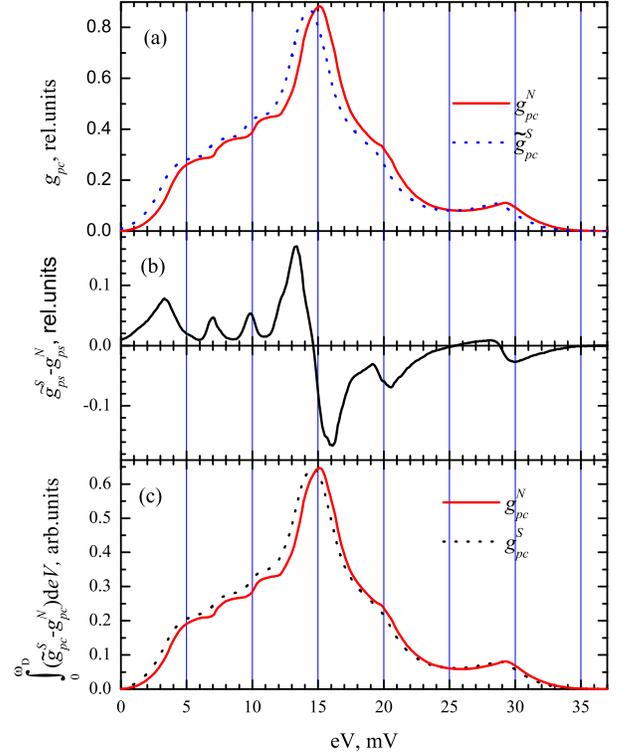}
\caption[]{${\rm g}_{{\rm PC}}^{{\rm N}} $ is the EPI function of Sn-Cu point
contact reconstructed from
the spectrum shown in Fig.\ref{Fig2} $\tilde{{\rm g}}_{{\rm PC}}^{{\rm S}} $
is the theoretically predicted point-contact EPI function upon transition into the
superconducting state (Eq.\eqref{eq__5}, see text for more detail) (a);
difference curve (b); integral of the difference curve, point-contact EPI function
obtained from the first derivative of the excess current
${\rm g}_{{\rm PC}}^{{\rm S}} $ in comparison with ${\rm g}_{{\rm PC}}^{{\rm N}} $.
For convenience of comparison, the maximum values of the curves are set the same.}
\label{Fig1}
\end{figure}
In comparison to the original $N$-curve, the $S$-curve
exhibits a shift of the EPI maxima towards lower energies by
the magnitude of the gap $\Delta $. Moreover, its amplitude is
somewhat smaller than the amplitude of the initial curve due
to an additional broadening by the smearing factor S, Eq.\eqref{eq__4}.
As already mentioned, in the superconducting state, the
EPI spectrum should appear in the first derivative of the
excess current. Indeed, if we subtract the initial N-curve
from the S-curve, we obtain the S-N curve:  $\tilde{g}_{pc}^{S} -g_{pc}^{N} $.
As follows from Eq.\eqref{eq__2}, the EPI function can be reconstructed
from the first derivative of the excess current:

\begin{equation}
\label{eq__6}
g_{pc}^{S} (eV)=\frac{1}{\Delta }\int\limits_0^{eV} {\left[
{\tilde{{g}}_{pc}^{S} (\omega )-g_{pc}^{N} (\omega )} \right]d\omega }
\end{equation}

It should be emphasized that $\tilde{g}_{pc}^{S} $ and $g_{pc}^{S} $
are different functions. The former one, given by Eq.\eqref{eq__5}, is proportional
to the second derivative of the CVC and reflects the transformation
of the spectrum (broadening and the shift of the phonon
peaks) upon transition of the heterojunction into the
superconducting state. The latter one, given by Eq.\eqref{eq__6}, see
also Eq.\eqref{eq__2}, is proportional to the first derivative of the
excess current and does not contain any additional broadening.
The position of the phonon maxima in the $g_{pc}^{S} (eV)$ is intermediate
between $\tilde{g}_{pc}^{S} $ and $g_{pc}^{N} $.
Note that for $S-c-S$ contacts,
the position of maxima in the EPI function reconstructed
from the first derivative of the excess current match that for
the normal condition.

\subsection{Elastic contribution}

The CVC of a point contact in which one or both electrodes
contain a superconductor with strong EPI comprises,
along with the above nonlinearities, an additional \textbf{elastic}
component of the current related to the frequency dependence
of the superconducting energy gap. This additional nonlinearity
arises due to the electron-phonon renormalization
of the energy spectrum of the superconductor and is manifested
as \textit{differential conductance maxima} in the region of
characteristic phonon energies in the first derivative of the
excess current, which are shifted to higher energies by the
magnitude of the superconducting energy gap \cite{Omel'yanchuk}.

Equation \eqref{eq__7}, which describes the first derivative of the
CVC in a point contact with direct conductivity, differs from
the corresponding expression \eqref{eq__8} for a tunnel junction,
\cite{Omel'yanchuk}

\begin{equation}
\label{eq__7}
\left( {\frac{dI}{dV}} \right)_{S-c-N} =\frac{1}{R_{0} }\left\{ {1+\left|
{\frac{\Delta \left( \varepsilon \right)}{\varepsilon +\sqrt {\varepsilon
^{2}-\Delta^{2}\left( \varepsilon \right)} }} \right|_{\varepsilon =eV}^{2}
} \right\}
\end{equation}

\begin{equation}
\label{eq__8}
\left( {\frac{dI}{dV}} \right)_{S-I-N} =\frac{1}{R_{0} }Re\left\{
{\frac{\varepsilon }{\sqrt {\varepsilon^{2}-\Delta^{2}\left( \varepsilon
\right)} }} \right\}_{\varepsilon =eV}
\end{equation}

This difference is due to Andreev reflection processes
leading to an excessive current in the region $eV~\gg~\Delta_0$.

Note that, unfortunately, the above equations do not
cover the most frequently encountered experimental situation-
point contacts with arbitrary transparency of the tunnel
barrier between the electrodes. In this respect, the
situation is similar to the attempts to determine the superconducting
energy gap prior to the BTK theory (Blonder,
Tinkham, and Klapwijk)\cite{Blonder}, which has provided a method for
determining the gap that takes into account an arbitrary barrier
transparency. It should be noted that for inelastic superconducting
spectroscopy this gap has been filled by Ref. \cite{Khlus3}.

Obviously, for point contacts with low barrier transparency,
it is the elastic contribution that is predominant due to
the suppression of the excess current. It has been noted in
Ref.\cite{Omel'yanchuk} that for ballistic contacts, CVC nonlinearities of elastic
origin may be comparable with the inelastic contributions
in point contacts. For the point contacts with direct conductivity
or high barrier transparency, the ratio between the elastic
and inelastic contributions is determined by the
parameters of the superconductor. As follows from Ref. \cite{Wolf},
the expected elastic contribution to the spectrum is proportional
to {$\sim{}$($T_C$/$\theta_D)^2$}, where $\theta_D$ is the Debye temperature.
Table~1 shows the elastic contributions for a number of
superconductors normalized by that of lead $\delta_{rel}$, which have
been studied in the previous publication\cite{Bobrov} and in the present
paper. The data from Ref.\cite{Wolf} were taken as a basis. We normalized
the data by the elastic contribution of lead since it
has the highest elastic contribution among the considered
superconductors. Table~1 also shows the energy gap and
superconducting transition temperature.

\begin{table}[]
\caption[]{Estimated elastic spectral contribution normalized by that of Pb, $\delta_{rel}$,
superconducting gap and transition temperature for several superconductors (SC).}
\small \begin{tabular}{|p{28pt}|l|l|l|l|l|l|l|}
\hline
SC & Pb & In & Sn & Ta & Al & NbSe$_2$ & MgB$_2$ \\
\hline
$\delta_{rel}$ & 1 & 0.21 & 0.078 & 0.063 & 0.00168 & 0.023 & 0.24 \\
\hline
$\Delta_0$,mV&
1.365&
0.525&
0.575&
0.7&
0.17&
1.07\textdiv2.48&
1.8\textdiv7.4 \\
\hline
T$_C$, K&
7.2&
3.415&
3.722&
4.47&
1.181&
7.2&
39 \\
\hline
\end{tabular}
\label{Table1}
\end{table}

Recall that for $S-c-N$ point contacts, the \textbf{inelastic} superconducting
contribution to the spectrum manifests itself as
\textit{differential \textbf{resistance} maxima} in the first derivative of the
excess current, which are shifted to lower energies by the
distance of the order of the gap. On the other hand, there is
no such shift for $S-c-S$ point contacts. Therefore, \textit{these contributions
oppose each other} and, if their magnitude is similar,
might attenuate each other. Since the inelastic
contribution is proportional to the magnitude of the excess
current, i.e., $\Delta $, and the elastic contribution is proportional to
($\Delta /E)^2$ (Ref.\cite{Wolf}) (see also Eq.\eqref{eq__7}),
starting from a certain value of $\Delta $, the elastic contribution dominates.

It can be expected that the positions of the maxima in
the EPI functions reconstructed from $S-c-S$ and $S-c-N$ point
contacts, as well as that for weakly coupled superconductors,
will be different.

Both for tunneling and point contacts with direct conductivity,
the elastic contribution to the spectrum does not
explicitly contain the EPI function g($\omega $). However, it can
be reconstructed by inverting the Eliashberg equations
(similar to the case of Rowell-McMillan's elastic tunneling
spectroscopy \cite{Rowell1}).

\section{Reconstruction of the EPI functions}

\subsection{Sn-based point contacts}

\begin{figure}[t]
\includegraphics[width=8cm,angle=0]{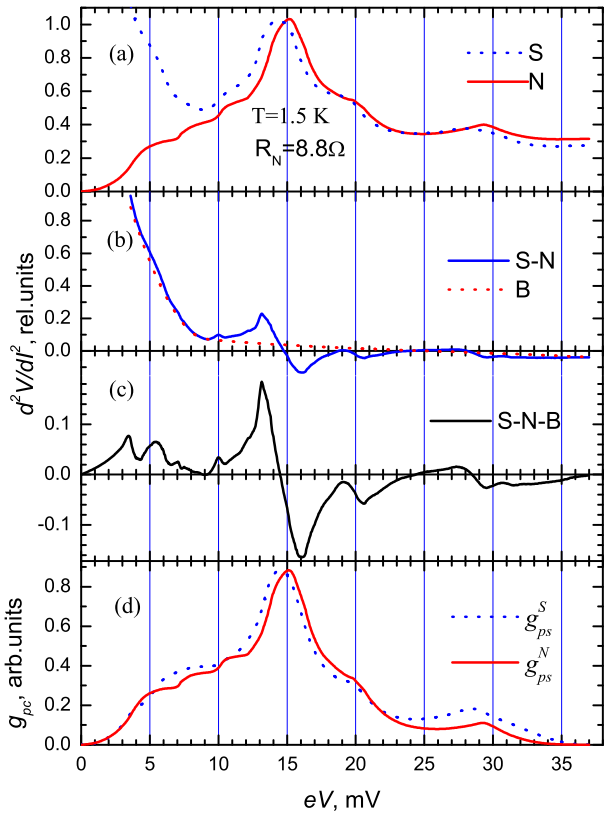}
\caption[] {EPI spectra of Sn-Cu point contact in the normal and superconducting
states. Superconductivity is suppressed by a magnetic field (a); the difference
between the spectra in the superconducting and normal states and
the estimated background curve (b); difference curve (after subtracting the
background B) (c); point-contact EPI function reconstructed by integrating
the curve in panel (c) versus the EPI function of the normal state (d).}
\label{Fig2}
\end{figure}

Fig. \ref{Fig2} (a) shows the spectra of Sn-Cu point contacts in
the normal and superconducting states.\cite{Yanson} Markedly lower
level of background in the superconducting spectrum and the
presence of the gap peak in the region of low energies
requires, similar to the previous work,\cite{Bobrov} that the background
$B$ is subtracted from the difference curve $S-N$. The difference
curve with the background subtracted, $S-N-B$, is very
close to the theoretically calculated curve $\tilde{g}_{pc}^{S} -g_{pc}^{N} $
in Fig. \ref{Fig1}. Finally, the lower part of the figure shows a comparison of
the PC EPI function reconstructed from the spectrum in the
normal state (curve $g_{pc}^{N} $) and the EPI function reconstructed
from the superconducting contribution to the spectrum
(curve~$g_{pc}^{S}$). For convenience of comparison, the curves are
plotted with equalized amplitude. There is excellent agreement
with the theoretically predicted behavior of the superconducting
EPI function--the shift of the maxima to lower
energies by the distance of the order of the gap. A slight mismatch
in the shape of the curve reconstructed from the experimental
data as compared to the calculated EPI function
in the region of large displacements can be related to certain
arbitrariness in defining the background or an increasing
contribution from the peripheral regions of the point contact.
The latter can occur due to the increasing concentration of
nonequilibrium phonons in these regions caused by decreasing
the electron energy relaxation length in the vicinity of
the Debye energy.

\begin{figure}[t]
\includegraphics[width=8cm,angle=0]{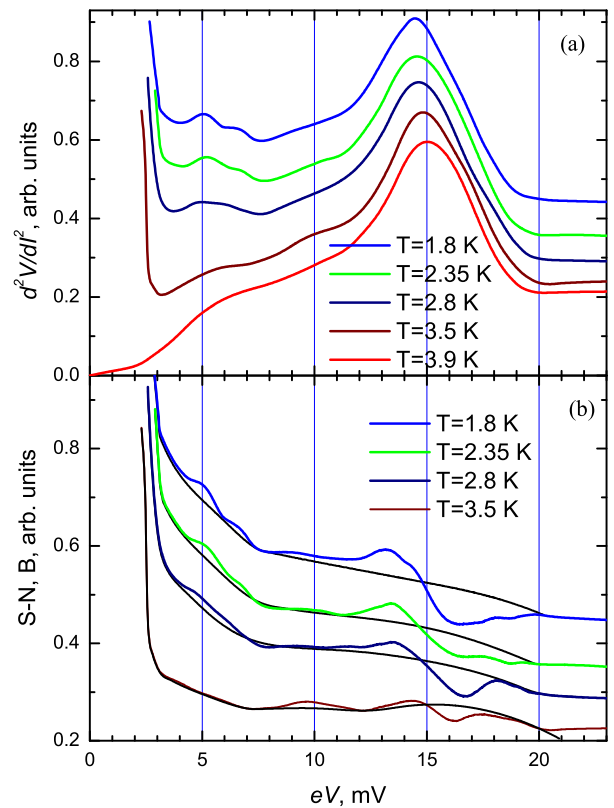}
\caption[] {Point-contact spectra of Sn in the normal and superconducting state,
adopted from Ref.\cite{Kamarchuk1}. $H = 0$ (a).
Superconducting contribution to the spectrum at different
temperatures and the estimated background curves (b).
$T = 1.8 K$: $T/T_C = 0.48$, $\Delta = 0.96\Delta_0$;
$T = 2.35 K$:  $T/T_C = 0.63$, $\Delta = 0.89\Delta_0$,
$T = 2.8 K$: $T/T_C = 0.75$, $\Delta = 0.78\Delta_0$,
$T = 3.5 K$: $T/T_C = 0.94$, $\Delta = 0.41\Delta_0$.}
\label{Fig3}
\end{figure}

As already noted, in the case of $S-c-S$ contact, the EPI
function reconstructed from the first derivative of the excess
current exhibits the same position of the maxima as the EPI
function of the normal state. Although an expression similar
to Eq.\eqref{eq__3} describing the transformation of the second derivative
of CVC upon the transition of electrodes into the $S$-state
has not been given in Ref.\cite{Khlus1}, from the similarity of the
expressions \eqref{eq__1} and \eqref{eq__2}, we can assume that the algorithm
used for $S-c-N$ point contacts can be employed here as well.
Fig.\ref{Fig3}(a) shows a set of the second derivatives of CVC
obtained in the normal and superconducting states,\cite{Kamarchuk1} and
Fig.\ref{Fig3}(b) displays the spectral contribution associated with
superconductivity as well as the estimated background
curves.

\begin{figure}[t]
\includegraphics[width=8cm,angle=0]{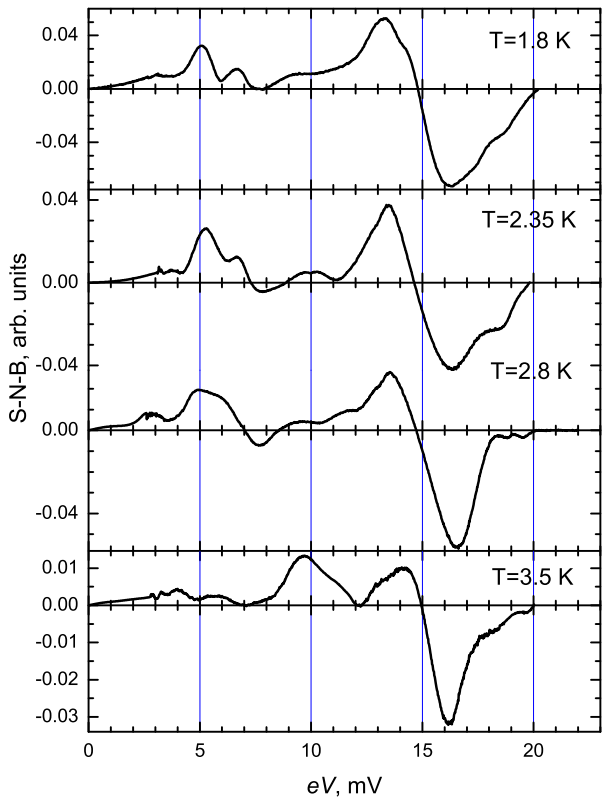}
\caption[] {Superconducting contribution to the point-contact spectra of Sn at
different temperatures after subtracting the background curves (Fig.\ref{Fig3}).}
\label{Fig4}
\end{figure}

\begin{figure}[t]
\includegraphics[width=8cm,angle=0]{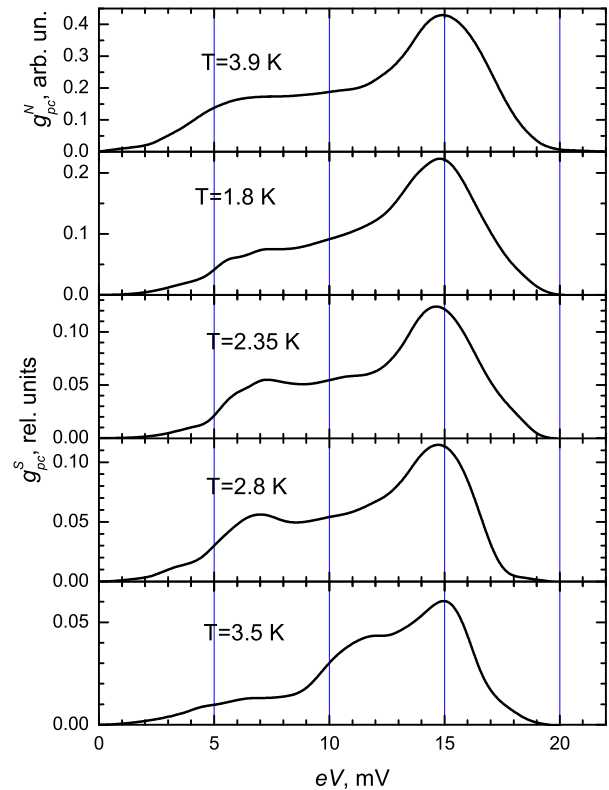}
\caption[] {Point-contact EPI functions of Sn, which were reconstructed from
the difference curves shown in Fig.\ref{Fig4}.}
\label{Fig5}
\end{figure}

Fig.\ref{Fig4} shows the difference curves after background
subtraction. Despite the fact that, unlike the previous case
for the Sn-Cu point contact, the temperatures of the normal
and superconducting states are not the same, the reconstruction
of the EPI functions from the superconducting spectral
contribution (Fig.\ref{Fig5}) is quite satisfactory.

Minor variations in
the shape of the curves can be easily explained by certain
arbitrariness in defining the background. Even for temperatures
near $T_C$, the reconstructed curve matches the normal
state spectrum quite satisfactory and the agreement can be
further improved by a better choice of the background curve.

\subsection{Al-based point contacts}

\begin{figure}[t]
\includegraphics[width=8cm,angle=0]{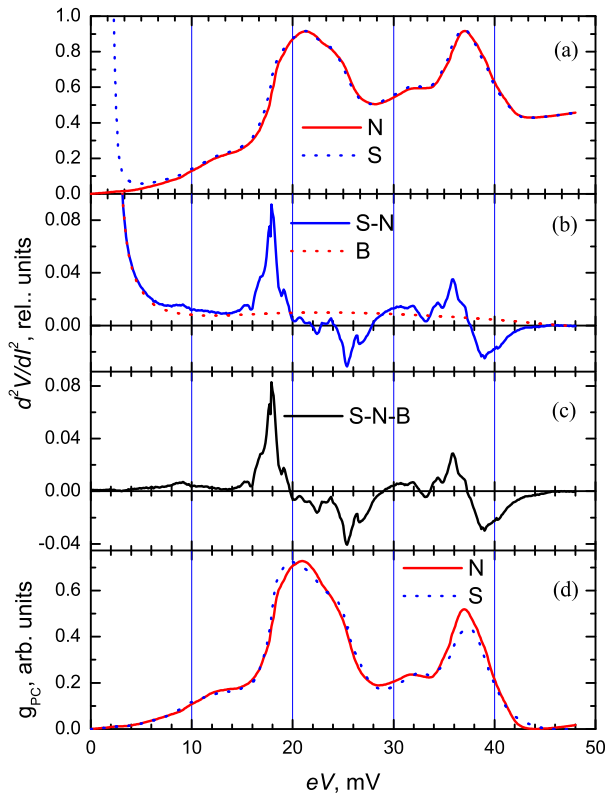}
\caption[] {EPI spectra of an Al-Al point contact in the normal and
superconducting states. $T/T_C = 0.68$, $\Delta = 0.85\Delta_0$.
Superconductivity is suppressed by a magnetic field (a).
The difference between the superconducting and normal
spectra and the estimated background curve (b). Difference curve (after
background subtraction) (c). Point-contact EPI function reconstructed by
integrating the difference curve in panel (c) versus the EPI function of the
normal condition (d).}
\label{Fig6}
\end{figure}

Aluminum has a relatively low superconducting transition
temperature and a small value of the superconducting
energy gap (Table~1). This means a small superconducting
contribution to the spectrum. Together with the inevitable
inaccuracies arising when scanned experimental curves are
digitized, this leads to a relatively low accuracy of the difference
curve obtained. Nevertheless, the curves shown in
Fig.\ref{Fig6} (similar to the curves published in Ref.\cite{Chubov})
demonstrate that the normal and superconducting EPI functions
match each other sufficiently well.

\subsection{Pb-based point contacts}

\begin{figure}[t]
\includegraphics[width=8cm,angle=0]{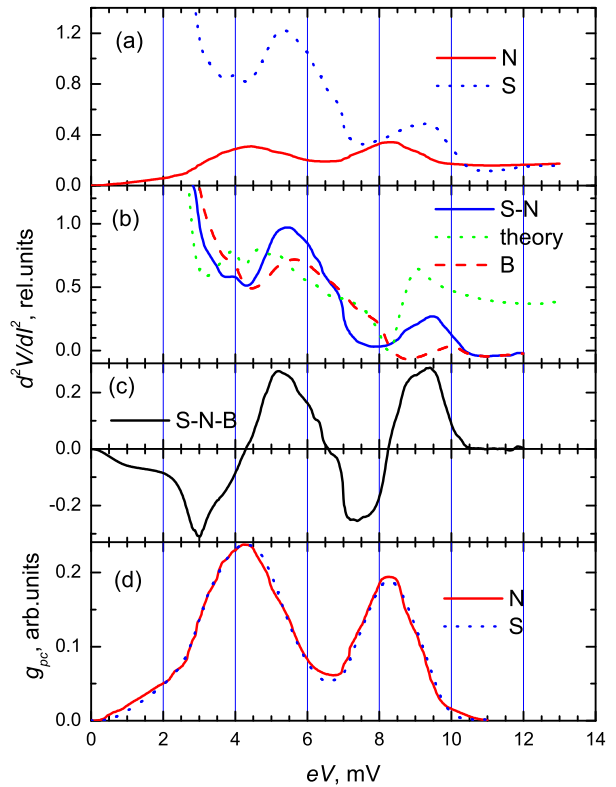}
\caption[] {EPI spectra of a Pb-Ru point contact in the normal and superconducting
states. Superconductivity is suppressed by a magnetic field (a). The
difference between the superconducting and normal spectra and the estimated
background curve. The dashed line shows the theoretically calculated
elastic contribution to the spectrum (see text) (b). Difference curve (after
background subtraction) (c). Point-contact EPI function reconstructed by
integrating the difference curve in panel (c) versus the EPI function of the
normal condition (d).}
\label{Fig7}
\end{figure}

Lead has highest elastic contribution of known superconductors
(Table~1). In Ref.\cite{Khotkevitch2}, the second derivatives of CVC
for Pb-Ru heterojunctions have been measured in both the
superconducting $S$ and normal $N$ states (Fig.\ref{Fig7}).

The EPI spectrum
of ruthenium does not overlap in energy with the spectrum
of lead and therefore was not taken into account. The
intensity of the reduced spectrum of lead is close to the maximum
for a symmetric heterojunction (0.4 of the maximum intensity
for a homojunction). Full contribution to the spectrum
associated with superconductivity (the difference curve $S-N$),
shown in Fig.\ref{Fig7}(b), is quite different from similar contributions
to the spectrum in metals with weak electron-phonon interaction
and does not allow to restore the EPI function using the
methods previously employed, in particular, subtraction of a
smooth background. The figure also shows the second derivative
$d^2V/dI^2(eV)$ of the elastic superconducting contribution to
the spectrum ("theory" curve), which was obtained from the
$dI/dV$ dependence found from Eq.\eqref{eq__7} by numerical differentiation.
When calculating $dI/dV$, the tables of real and imaginary
parts of $\Delta $($\varepsilon $) \cite{Rowell2} obtained from
tunneling experiments were used.

Although the calculated second derivative of the elastic
contribution ("theory" curve) is similar to the difference curve
$S-N$, there are notable differences, especially at high energies.
As already mentioned, the elastic superconducting contribution
manifests itself as maxima of the differential
conductance in the region of characteristic phonon energies in
the first derivative of the excess current. However, the difference
curve $S-N$ contains not only the elastic contribution, but
also inelastic one, and, apparently, in the same way as in the
superconductors with weak coupling, also additional nonlinearity
which is not accounted by the theory and is what we
call a superconducting background. Therefore, to obtain the
EPI spectrum by integration, as was done previously, let us
try to subtract the background from the difference curve $S-N$
using the same rules as in the case of weakly coupled superconductors.
\textit{After background subtraction, the areas under the
curve above and below the abscissa should be the same; for
energies above the Debye energy, the curve obtained after
background subtraction must be zero.}

The obtained background B is shown in Fig.\ref{Fig7}(b) as a
dashed curve, and the resulting curve after background subtraction,
$S-N-B$ is displayed in Fig.\ref{Fig7}(c). Fig.\ref{Fig7}(d) shows the
EPI $N$ function reconstructed from the spectrum in the normal
state and the EPI $S$ function obtained by integrating the curve
$S-N-B$. Here it is necessary to emphasize the following
points. First, as follows from the theoretical predictions, the
maxima of the function correspond to the maxima of the differential
conductance and not to the maxima of resistance as
in the case of the superconductors with weak EPI. After integration
the curve is inverted. Secondly, the positions of the
phonon peaks in the both curves coincide, and there is no shift
of the phonon peaks in the restored EPI function for S-c-N
point contacts. And finally, the background curve is not
smooth and monotonic but is similar in shape to the theoretically
calculated elastic contribution marked as "theory". Note
that for $S-c-S$ contacts, as discussed below, there is a shift of
the phonon maxima in the reconstructed EPI function to
higher energies by the distance of the order of $\Delta $.

\begin{figure}[t]
\includegraphics[width=8cm,angle=0]{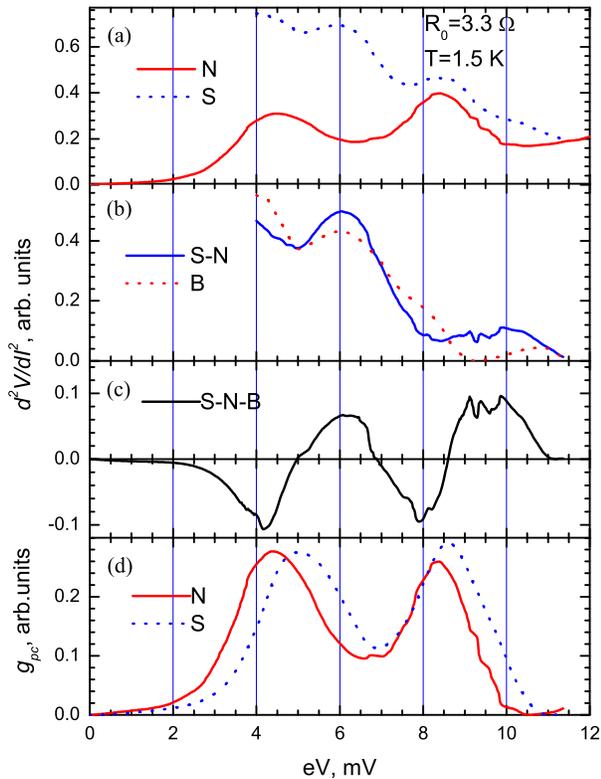}
\caption[] {EPI spectra of a Pb-Pb point contact in the normal and superconducting
states. Superconductivity is suppressed by a magnetic field (a). The
difference between the superconducting and normal spectra and the estimated
background curve (b). Difference curve (after background subtraction)
(c). Point-contact EPI function recovered by integrating the curve in
panel (c) versus the EPI function of the normal state (g).}
\label{Fig8}
\end{figure}

Fig.\ref{Fig8} shows the spectra of a Pb-Pb point contact in the
normal and superconducting states \cite{Kamarchuk2}.
The curves were treated similar to the previous case. As anticipated above,
the position of the phonon peaks in the EPI function reconstructed
from a $S-c-S$ contact differs from that of a $S-c-N$
contact and is shifted toward higher energies by a distance of
the order of $\Delta $.

\begin{figure}[t]
\includegraphics[width=8cm,angle=0]{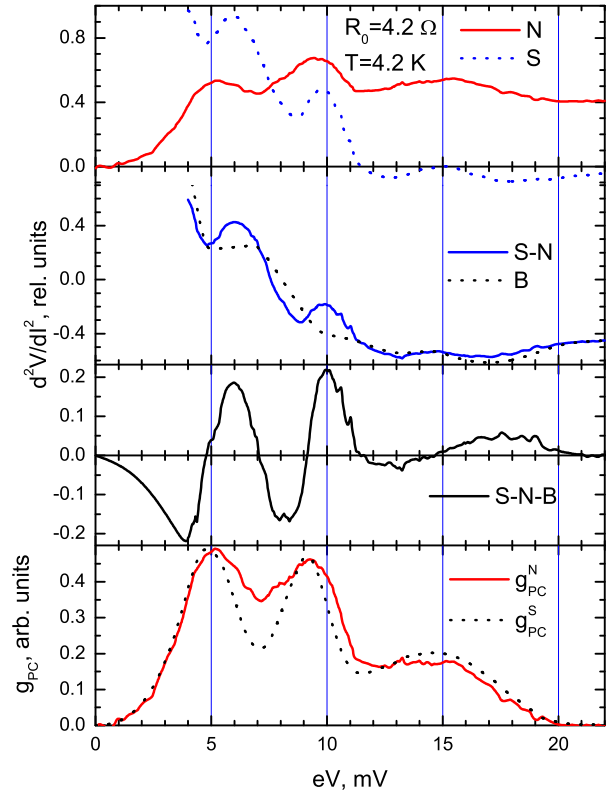}
\caption[] {EPI spectra of a $Pb-Sn$ point contact in the normal and superconducting
states. Measurement temperature is above $T_C$ of Sn
($T/T_C = 0.58$, $\Delta = 0.92\Delta_0$).
Superconductivity is suppressed by a magnetic field (a). The difference
between the superconducting and the normal spectra and the estimated
background curve (b). Difference curve (after background
subtraction) (c). Point-contact EPI function reconstructed through integrating
the difference curve in panel (c) versus the EPI function of the normal
state (d).}
\label{Fig9}
\end{figure}

Finally, the data for a Pb-Sn point contact at temperatures
above $T_C$ for Sn are shown in Fig.\ref{Fig9}.\cite{Kamarchuk2}
In this case, the tin contribution is considerably higher than that of ruthenium,
and it overlaps in energy with the spectrum of lead.
Thus it should be necessarily taken into account in data processing.
Since this is an $S-c-N$ point contact, there is no shift
of the phonon peaks to higher energies in the EPI function
reconstructed from the superconducting contribution.

\subsection{In-based point contact}

\begin{figure}[t]
\includegraphics[width=8cm,angle=0]{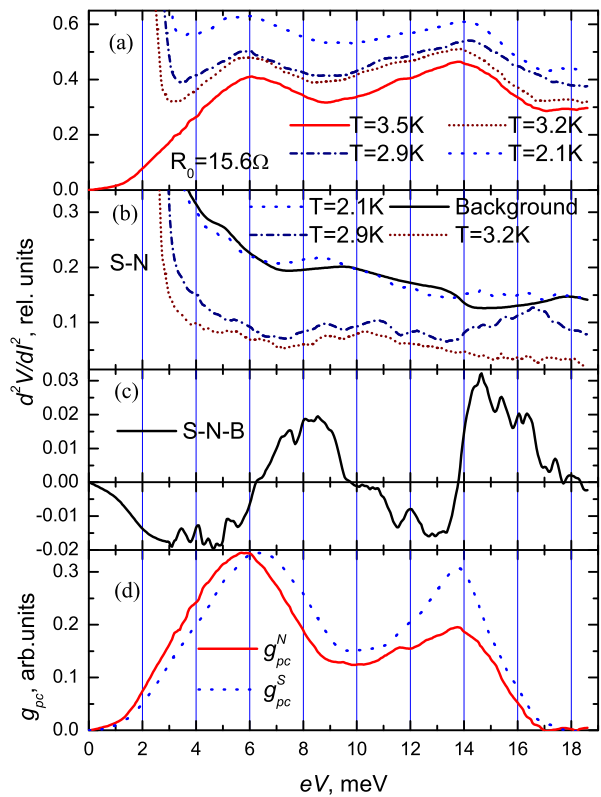}
\caption[] {EPI spectra of an In-In point contact in the normal and superconducting
states at different temperatures (a). The differences between the
superconducting and normal spectra, as well as the estimated background curve.
$T = 2.1 K$: $T/T_C = 0.62$, $\Delta = 0.89\Delta_0$;
$T = 2.9 K$: $T/T_C = 0.85$, $\Delta = 0.63\Delta_0$;
$T = 3.2 K$: $T/T_C = 0.94$, $\Delta = 0.41\Delta_0$ (b).
Difference curve (after background subtraction) (c). Point-contact EPI function
reconstructed by integrating the difference
curve (c) versus the EPI function of the normal state (d).}
\label{Fig10}
\end{figure}

In the case of lead, the superconducting contribution to
the spectrum associated with EPI is very large and, as can be
seen in Fig.\ref{Fig7}, its amplitude exceeds non-linearity in the
normal state. Indium is intermediate in EPI strength and, as
follows from the table, exhibits a five-fold smaller elastic
contribution to the spectrum compared to lead, but 2.7-fold
higher than tin. At the same time, the superconducting transition
temperature and the gap are only slightly ($\sim8\%$) less
than those of tin, so the inelastic contribution to the spectrum
must be very close for these metals. Since, as noted above,
the elastic and inelastic contributions counteract each other,
in the case of indium they should, to a large extent, weaken
each other. Fig.\ref{Fig10} shows the spectra of indium\cite{Kamarchuk1}.
As can be seen in the figure, the contribution to the spectrum associated
with superconductivity in indium is very small and of elastic
nature. Unlike tin, where it was possible to restore the spectrum
from the superconducting contribution in a wide temperature
range (Figs.\ref{Fig3} and \ref{Fig4}), for indium, this was possible
only at the lowest temperature.

\section{Discussion}

In determining the function of electron-phonon interaction,
the traditional tunneling spectroscopy is limited to the
superconductors with strong coupling. At the same time
Yanson's point-contact spectroscopy focusses on metals in
the normal state. Inelastic superconducting point-contact
spectroscopy thus fills the gap and can be used to study
superconductors with weak coupling. Moreover, as follows
from the discussion of indium, the superconductors in which
the elastic and inelastic contributions are close in value are
the most complex objects since the elastic and inelastic
contributions counteract each other, which leads to a weakening
of the resulting contribution to the spectrum. When
evaluating the sign and magnitude of the expected effect in
such superconductors, first of all the transparency of the barrier
between the electrodes should be taken into account. A
situation is very likely to appear in which contacts with the
different transparency of barriers exhibit different positions
of the spectral maxima due to the predominance of elastic or
inelastic spectral contributions. Note that it is not so much
the magnitude of the elastic contribution to the spectrum as
the ratio between the elastic and inelastic contributions that
is important. For instance, as follows from the data in Table~1,
the elastic contribution to the spectrum in MgB$_2$ is slightly
larger than that in In. However, the resulting spectrum in
MgB$_2$ is inelastic\cite{Bobrov}. Here we should give attention to the
magnitude of the superconducting gaps in In and MgB$_2$. The
inelastic contribution is proportional to the energy gap,
which in MgB$_2$ is an order of magnitude larger than in In.

Even for superconductors with weak coupling there may
be deviations from the theoretical predictions. For instance
(see Table~1), the elastic contribution to the spectrum in Sn
is greater than that in Ta by $\sim12\%$. However, due to a
shorter electron energy relaxation length, the presence of
phonons with low group velocity results in a significant
influence of the contact region on the formation of the superconducting
contribution. As a result, sharpening of the
phonon peaks and other deviations from the theoretical predictions
are observed. Another typical example is 2H-NbSe$_2$,
a superconductor with covalent bonds between the atoms
within the layer and van der Waals forces between the
layers. Therefore, both the current spreading and dispersion
of phonons in the point contacts based on NbSe$_2$ are anisotropic.
This leads to a slower decrease of the concentration
of nonequilibrium phonons with increasing the distance
from the constriction, and thus, an increase of the superconducting
contribution to the spectrum. As can be seen in Fig.~3
in Ref.\cite{Bobrov}, this contribution is sufficiently large and only an
order of magnitude smaller than the gap peculiarity in the
spectrum.

To summarize, let us note that all of the observed deviations
from the theoretical predictions are, in varying degrees,
related to the influence of the contact region that requires further
theoretical and experimental studies. Moreover, point contacts
with a non-uniform distribution of impurities exhibiting
diffusion transport through the junction and ballistic banks
also require further studies. It should be noted that this situation
is most easily achieved for superconductors with covalent
bonds between atoms which have a rigid crystal lattice. For ordinary
metals, due to their ductility, the lattice distortions upon
forming point contacts could extend to the banks as well.

\section{Conclusions}

\begin{enumerate}

\item The EPI functions for Sn and Al were reconstructed from
the superconductivity-related contributions to the spectra of
point contacts based on these metals. The procedure of
reconstruction of the EPI functions produces similar results
across a wide range of temperatures. As follows from the
theory of superconductors with weak coupling, the superconducting
inelastic contribution to the spectrum manifests
itself as \textit{differential resistance maxima} in the first derivative
of the excess current in the range of characteristic phonon
energies. The position of these peaks coincides with the
phonon peaks observed in the normal state of $S-c-S$ point
contacts and, for $S-c-N$ point contacts, is shifted to \textit{lower}
energies by the value of the superconducting energy gap.

\item The EPI functions for Pb and In were reconstructed from
the superconductivity-related contributions to the spectra
of point contacts based on these metals. For superconductors
with strong coupling, the superconducting elastic
contribution to the spectrum manifests itself as \textit{differential
conductivity maxima} in the first derivative of the excess
current in the range of characteristic phonon energies.
The position of these peaks \textit{coincides} with the phonon
peaks observed in the normal state of $S-c-N$ point contacts
and, for $S-c-S$ point contacts, is \textit{shifted to higher energies}
by the value of the superconducting energy gap.

\end{enumerate}

\end{document}